\documentclass{aa}
\usepackage{psfig}

\begin{document}

\title{Modelling the rotational curves of spiral galaxies with a scalar field}

\author{J.P. Mbelek}
\institute{Service d'Astrophysique,
C.E. Saclay,  F-91191 Gif-sur-Yvette Cedex, France}
\authorrunning{Mbelek}
\titlerunning{A  model of rotational curves}

\date{Received / Accepted }

\abstract{In a previous work (Mbelek 2001), we modelled the
rotation curves (RC) of spiral galaxies by including in the equation of motion
of the stars the dynamical terms from an external real self-interacting scalar
field, $\psi$, minimally coupled to gravity and which respects the equivalence
principle in the weak fields and low velocity approximation.  This model
appeared to have three free parameters :  the turnover radius, $r_{0}$, the
maximum tangential velocity, $v_{\theta max} = v_{\theta}(r_{0})$, plus a
strictly positive integer, $n$.  Here, we propose a new improved version where
the coupling of the $\psi$-field to dark matter is emphasized at the expense of
its self-interaction.  This reformulation presents the very advantageous
possibility that the same potential is used for all galaxies.  Using at the same time a quasi-isothermal dark matter density and the scalar
field helps to better fit the RC of spiral galaxies.  In addition, new
correlations are established.}

\maketitle

%

\section{Introduction}\label{intro}

Since 1998, many groups have been working on modeling the RC of
spiral galaxies with ultra-light scalar fields as an alternative to the standard
assumption of weakly-interacting massive particles.  Guzman and Matos (2000)
presented a model where they assume that the galactic dark matter consists
entirely of a self-interacting real scalar field $\Phi$ endowed with a potential
$V(\Phi)$ which is an exponential function of $\Phi$.  First they obtained an
exact static and axially symmetric solution of the Einstein-dilaton
theory (Guzman \& Matos 2000a).  This solution yields an effective energy density
$\mu_{DM} = V(\Phi)$ which mimics an isothermal dark halo with a constant core
density.  The resulting circular velocity profile of test particles involves
three free parameters to fit the observations.  However, the authors (Guzman \& Matos 2000b) pointed out
some difficulties related to the physical origin of the kind of exponential
potential $V(\Phi)$ they have to consider.  Based on a complex
scalar field forming galactic halos after Bose condensation, Arbey {\sl et al.} (Arbey et al. 2001a)
have suggested a massive ($m = 0.4$ to $1.6$~$10^{-23}$ eV/$c^{2}$) and
non-interacting complex scalar field, as a dark matter candidate alternative to
the neutralino, to bind galaxies and flatten the RC of spirals.
Their motivation for a complex ({\sl i.  e.}, charged) scalar field follows from
the need for stable bounded configurations in the case of self-interacting bosons
(not interacting with any other kind of matter).  The model of the authors
accounts better for the dark matter inside low-luminosity spirals than for the
brightest objects where baryons dominate.  However, because they deal with a
charged scalar field (Arbey et al. 2001b), some problems (pointed out by the authors themselves)
arise when the limits from galactic dynamics and some cosmological constraints
are taken simultaneously into account.  Ure\~na-L\`opez
and Liddle (Ure\~na-L\`opez \& Liddle 2002) have shown that the supermassive black holes hosted at the center of
galaxies can coexist with scalar field halos.  Since all the
aforementioned studies consider only self-interacting scalar fields, the
equation satisfied by their respective scalar field is of the sourceless
Klein-Gordon type (including the derivative of the scalar field potential).  The
aim of this work is to show that a real scalar field, $\psi$, minimally coupled
to gravity sourced by and interacting with matter (both baryonic and dark
matter), also may reproduce the RC of spiral galaxies.  This can be achieved if
the force term involved by the scalar field can contribute sufficiently to lower
the magnitude of the orbital momenta of stars and gas, even if $\psi$ itself
does not contribute significantly to the mass of the dark halo.  The general
equation of $\psi$ reads (Mbelek \& Lachi\`eze-Rey 2003) \begin{equation} \label{scalar field GR
eq} {\nabla}_{\nu} {\nabla}^{\nu} {\psi} = \,- \,J \,- \,\frac{\partial J} {\partial {\psi}} \,\psi \,-
\,\frac{\partial U} {\partial {\psi}}, \end{equation} where $U = U(\psi)$ and $J
= J(x^{\alpha})$ denote respectively the self-interaction potential and the
source term of the $\psi$-field ; ${\nabla}_{\nu} {\nabla}^{\nu}$ is the d'Alembertian of the curved
four dimensional spacetime endowed with the metric ${ds}^{2} = g_{\mu\nu}
\,dx^{\mu} \,dx^{\nu}$.  The potential $U$ is of a symmetry breaking type but
needs not to be specified here ; $U({\psi}_{\infty}) = 0$, $\frac{\partial
U}{\partial \psi} ({\psi}_{\infty}) = 0$ and $\frac{\partial^{2} U}{\partial
\psi^{2}} ({\psi}_{\infty}) > 0$, where ${\psi}_{\infty}$ denotes the value of
$\psi$ at the minimum of the potential $U(\psi)$.  Equation (\ref{scalar field
GR eq}) describes a minimally coupled classical real scalar field with the usual
Lagrangian density (Narlikar \& Padmanabhan 1986) \begin{equation} \label{Lagrangian of psi}
{\mathcal L} = \frac{c^{4}}{8\pi G} \,[ \,\frac{1}{2} \,{\partial}_{\mu} \psi
\,{\partial}^{\mu} \psi \,- \,U \,- \,J \psi \,] \,\sqrt{-\det{(g_{\mu\nu})}},
\end{equation} but taking into account that the source term, $J$, may depend on
the $\psi$-field too.  This is the case when the coupling constants to the other
form of matter-energy are generalized to be functions that depend also on the
scalar field.  For our purpose, one may neglect the contribution of the
electromagnetic fields.  Hence, $J$ accounts for the matter source only.  As
usual in the framework of scalar-tensor type theories, this is of the general
form $J = \frac{4\pi G}{c^{4}} \,g(\psi) \,T^{\alpha}_{\alpha}$ (see {\sl e.
g.}, Straumann 2000), where $T^{\alpha}_{\alpha}$ is the trace of the
energy-momentum tensor of the matter source and the coupling function $g$ is
such that $g({\psi}_{\infty}) = 0$ in the vacuum state (minimum of the potential
$U$).

\section{Weak and static field approximation} Assuming that the excitation
$\delta \psi = \psi \,- \,{\psi}_{\infty}$ of the $\psi$-field is small compared
to ${\psi}_{\infty}$, the equation of the $\psi$-field reads in the first
approximation for a static matter distribution \begin{equation} \label{eq psi}
\Delta \delta \psi \,- \,{\lambda}^{-2} \,\delta \psi = \frac{\partial g}
{\partial {\psi}} ({\psi}_{\infty}) \,{\psi}_{\infty} \,\frac{4\pi G}{c^{2}}
\,\rho, \end{equation} where $\lambda = \left( \,\frac{\partial^{2} U} {\partial
{\psi}^{2}} ({\psi}_{\infty}) \,\right)^{-1/2}$ denotes the range of the
$\psi$-field and $\rho$ is the mass density of the matter fields other than the
$\psi$-field itself.  Out of the matter source, eq(\ref{eq psi}) reduces to
\begin{equation} \label{eq psi vacuum sol.}  \Delta \delta \psi \,-
\,{\lambda}^{-2} \,\delta \psi = 0, \end{equation} whose solution is the
familiar Yukawa potential $\delta \psi \propto \exp{( \,-\,\lambda r \,)}/r$.
Accordingly, $\psi \rightarrow {\psi}_{\infty}$ at infinity.  Since $\psi$ is
long range (say $\lambda \geq 100$ kpc), we may neglect in the first
approximation the scalar field mass term, ${\lambda}^{-2}$, with respect to the
mass density terms within galaxies.  Thus, eq(\ref{eq psi}) simplifies as
follows \begin{equation} \label{eq psi bis} \Delta \psi = \frac{\partial g}
{\partial {\psi}} ({\psi}_{\infty}) \,{\psi}_{\infty} \,\frac{4\pi G}{c^{2}}
\,\rho.  \end{equation}

\section{The equation of motion} The equation of motion of test bodies is
obtained following the usual procedure (Ciufolini \& Wheeler 1995) which consists of
considering a gas of dust (pressureless perfect fluid) whose energy-momentum tensor
reads $T^{\mu\nu} = \rho \,c^{2} \,u^{\mu} \,u^{\nu}$, where the density $\rho$
is a constant.  The equation of motion is then derived from the conservation of
the energy-momentum tensor, that is ${\nabla}_{\nu} \,T^{\mu\nu} = 0$ (where
${\nabla}_{\nu}$ denotes the covariant derivative ${\nabla}_{\nu} T^{\mu\nu} =
{\partial}_{\nu} T^{\mu\nu} \,+
\,{\Gamma}^{\sigma}_{\sigma\nu} \,T^{\mu\nu} \,+
\,{\Gamma}^{\mu}_{\sigma\nu} \,T^{\sigma\nu}$, and ${\Gamma}^{\mu}_{\sigma\nu} =
\frac{1}{2} \,g^{\mu\lambda} \,( \,\frac{\partial g_{\sigma\lambda}}{\partial
x^{\nu}} \,+ \,\frac{\partial g_{\lambda\nu}}{\partial x^{\sigma}}
\,-\,\frac{\partial g_{\sigma\nu}}{\partial x^{\lambda}} \,)$ defines the
Christoffel symbols).  In our framework, the same procedure applies to an
appropriate effective energy-momentum tensor $T_{eff}^{\mu\nu}$ which accounts
for the presence of the scalar field.  Since the trace $T^{\alpha}_{\alpha}$ is
involved in the source term $J$, arranging in order to isolate all the
contributions of the matter in the total Lagrangian density leads to the
identification $T_{eff}^{\mu\nu} = T^{\mu\nu} \,+ \,g(\psi) \,\psi
\,T^{\alpha}_{\alpha} \,g^{\mu\nu}$.  As a consequence, this does not lead to
the geodesic equation since the influence of the scalar field now enters.
Indeed, it follows for a neutral test body \begin{equation} \label{eq motion}
\frac{du^{\mu}}{ds} \,+ \,{\Gamma}^{\mu}_{\alpha\beta} \,u^{\alpha} \,u^{\beta}
= \frac{d( \,g(\psi)\,\psi \,)}{ds} \,\,u^{\mu} \,- \,{\partial}^{\mu} (
\,g(\psi)\,\psi \,).  \end{equation} Note that eq(\ref{eq motion}) is generic
in scalar-tensor theory like the Brans-Dicke (Brans \& Dicke 1961) theory or the 5D
compactified Kaluza-Klein theory (Wesson \& Ponce de Leon 1995).  Independently
of the physical meaning, eq(\ref{eq motion}) may be obtained from the
variational principle $0 = \delta \int m c \,\sqrt{g_{\mu\nu} \,u^{\mu}
\,u^{\nu}} \,ds$ with an effective mass $m \propto \exp{( g(\psi)\,\psi )}$.
The geodesic equation is recovered in the case of a non variable $\psi$-field.
Now, in the weak field and low velocity limit, one gets $\Gamma_{00}^{\mu}
\approx \,- \,\frac{1}{2} \,g^{\mu\mu} \,{\partial}^{\mu} g_{00}$ (with no
summation on $\mu$).  Hence, in the first order approximation, eq(\ref{eq
motion}) simplifies to \begin{equation} \label{EP dyn eq approxim vectorial}
\frac{d\vec{v}} {dt} = \,- \,\vec{\nabla} \left( \,V_{N} \,+ \,c^{2}
\,\frac{\partial g}{\partial \psi} ({\psi}_{\infty}) \,{\psi}_{\infty} \,\psi
\,\right) \,+ \,\frac{\partial g}{\partial \psi} ({\psi}_{\infty})
\,{\psi}_{\infty} \,\frac{d\psi} {dt} \,\vec{v}, \end{equation} where the
Newtonian potential, $V_{N}$, results from the contributions of the visible and
dark matter components including that of the $\psi$-field.  Newtonian physics is
recovered in the limit case $\delta \psi \rightarrow 0$.

\section{Modelling the RC of spirals} Since a drag-like force term appears in
eq(\ref{EP dyn eq approxim vectorial}), the rotational velocity $v$ should read
in the general form $v = \sqrt{v_{\theta}^{2} \,+ \,\dot{r}^{2}}$, where the
tangential velocity $v_{\theta}$ and the radial velocity $\dot{r}$ are derived
in spherical coordinates respectively from the tangential equation
\begin{equation} \label{EP dyn eq approxim vectorial ter} \frac{1}{r}
\,\frac{d(r\,v_{\theta})} {dt} = \frac{\partial g}{\partial \psi}
({\psi}_{\infty}) \,{\psi}_{\infty} \,\frac{d\psi} {dt} \,v_{\theta}
\end{equation} and the radial equation \begin{equation} \label{EP dyn eq
approxim vectorial quatro} \ddot{r} \,- \,\frac{v_{\theta}^{2}}{r} = \,- \,G
\,\frac{m(r) \,+ \,m^{(\psi)}(r)}{r^{2}}, \end{equation} where we have set
$m^{(\psi)}(r) = \frac{c^{2}}{G}\,\frac{\partial g}{\partial \psi}
({\psi}_{\infty}) \,{\psi}_{\infty}\,\frac{d\psi}{dr}\,r^{2}$ and $m(r) =
m_{vis}(r) \,+ \,m_{dark}(r)$ is the total mass up to radius $r$ assuming
spherical symmetry.  In the latter relation, $m_{vis}(r)$ denotes the visible
matter mass and $m_{dark}(r)$ the dark matter mass including both the
contributions of the dark halo, $m_{halo}(r)$, and that of the $\psi$-field.
So, integrating equation (\ref{EP dyn eq approxim vectorial ter}) above yields
\begin{equation} \label{angular momentum vs phi} \ln{(r\,v_{\theta})} =
\ln{{\mathcal J}} \,+ \,\frac{\partial g}{\partial \psi} ({\psi}_{\infty})
\,{\psi}_{\infty} \,( \,\psi \,- \,{\psi}_{\infty} \,), \end{equation} where
${\mathcal J}$ would represent the angular momentum per unit mass if the
$\psi$-field were not excited.  Hereafter, we assume $\frac{\partial g}
{\partial {\psi}} ({\psi}_{\infty})\,{\psi}_{\infty} > 0$ so that
$m^{(\psi)}(r)$ mimics a hidden mass profile\footnote{Since $\psi$ also acts as
a stabilizing field (see Mbelek \& Lachi\`eze-Rey 2003), one may expect that it affects the
dynamics within the galaxy by lowering the magnitude of the orbital momentum per
unit mass at any given radius.  This reads $r\,v_{\theta} < {\mathcal J}$ and
consequently $\delta \psi < 0$ on account of (\ref{angular momentum vs phi}) and
by assuming $\frac{\partial g}{\partial \psi} ({\psi}_{\infty})
\,{\psi}_{\infty}> 0$.  Consequently, the consistency with ${\psi}_{\infty}$
defining the minimum of the potential $U(\psi)$ at $r \rightarrow \infty$
implies that $d\psi/dr > 0$ for $\delta \psi$ decreasing monotonously.}.  In
addition, we adopt the model of a quasi-isothermal spherical dark halo, that is,
a mass density profile such that \begin{equation} \label{halo density}
{\rho}_{halo}(r) = {\rho}_{halo}(r_{0}) \,( \,\frac{r_{0}}{r} \,)^{2 +
\varepsilon} \end{equation} where $\varepsilon$ is a free parameter which
depends on the given galaxy and lies within the range $] 0, \frac{1}{2} ]$.  We
show below that the use of an isothermal dark halo together with the possibility
of non-circular orbits may explain the apparent discrepancy with the
Navarro-Frenk-White (NFW) or Burkert halos assuming circular orbits for the
latter.  Recently, Bournaud {\sl et al.} (Bournaud et al. 2003) have proved for the
first time from $N-$body simulation that a large extent of dark matter is
necessary for the formation of tidal dwarf galaxies.  In this study, the authors
described the dark matter as a standard isothermal sphere.  Hence, although
the main concern of the authors is the tidal interaction in the outskirts of
galaxies, the hypothesis of an isothermal (or quasi-isothermal sphere) still
proves to be worth investigating.  For simplicity, we will neglect the
contributions of the bulge and the disk (stellar plus gaseous disks) as sources
of the $\psi$-field since the dark halo is the major mass component of the
galaxies.  Thus, one finds that the static spherical solution $\psi$ of
eq(\ref{eq psi bis}) reads \begin{equation} \label{sol.  eq psi bis} \psi =
{\psi}_{\infty} \,- \,\frac{1}{\varepsilon} \,\frac{\partial g}{\partial \psi}
({\psi}_{\infty}) \,{\psi}_{\infty} \,\frac{G m_{halo}(r_{0})}{r_{0}\,c^{2}}
\,\left( \,\frac{r_{0}}{r} \,\right)^{\varepsilon}.  \end{equation} Inserting
the solution found for $\psi$ in relation (\ref{angular momentum vs phi}) above,
the tangential velocity expresses in the form \begin{equation} \label{velocity
vs radius} v_{\theta} = v_{\theta max} \,G_{\nu}(x), \end{equation} where $x =
r/r_{0}$ and we have set $\nu = 1/\varepsilon$ for convenience (see below).  The
functions $G_{\nu}$ are defined for $x > 0$ as follows \begin{equation}
\label{universal RC} G_{\nu}(x) = \frac{1}{x} \,\exp{[ \,\nu(1 - x^{-1/\nu})
\,]}.  \end{equation} These functions are positively defined, bounded from above
by $G_{\nu}(1) = 1$ and satisfy $G_{\nu}(x)^{k} = G_{k\nu}(x^{k})$.  Solving the
radial equation (\ref{EP dyn eq approxim vectorial quatro}), on account that
$\ddot{r} = d\dot{r}/dt = d(\frac{1}{2} \dot{r}^{2})/dr$ ($v_{\theta} =
v_{\theta}(r)$ and $\dot{r} = \dot{r}(r)$ at radius $r$), one finds
\begin{equation} \label{radial velocity} \dot{r}(r)^{2} = {\dot{r}}(r_{0})^{2}
\,+ \,2 \,\int_{r_{0}}^{r} \left( \,\frac{v_{\theta}^{2}}{r} \,- \,G
\,\frac{m(r) \,+ \,m^{(\psi)}(r)}{r^{2}} \,\right) \,dr.  \end{equation} Hence,
one expects $m(r) \,+ \,m^{(\psi)}(r) \simeq v_{\theta}^{2}\,r/G$ for any good
fit to the RC that can be obtained assuming circular orbits, where\footnote{The case $\varepsilon = 1$, which matches a NFW dark halo
density mass profile far beyond the core radius, yields a point mass $m^{(\psi)}
= \left( \frac{\partial g}{\partial \psi} ({\psi}_{\infty}) \,{\psi}_{\infty}
\right)^{2} \,m_{halo}(r_{0})$ that mimics a black hole located at the center of
the galaxy.}  $m^{(\psi)}(r) = \left( \frac{\partial g}{\partial \psi}
({\psi}_{\infty}) \,{\psi}_{\infty} \right)^{2} \,m_{halo}(r_{0}) \,(r/r_{0})^{1
- \varepsilon}$.  Actually, the departure from circular orbits may help to
understand the use of the various dark halo density mass profiles,
\begin{equation} \label{generic DM profile} \rho_{DM} = {\rho}_{*}
\,\frac{\left( \,\frac{r}{r_{*}} \,\right)^{\alpha\beta}}{\left(
\,\frac{r}{r_{*}} \,\right)^{2} \,( \,1 \,+ \,\frac{r}{r_{*}} \,) \,\left[ \,1
\,+ \,\left( \,\frac{r}{r_{*}} \,\right)^{\alpha} \,\right]^{\beta}},
\end{equation} that are advocated in the literature to fit the RC of spiral
galaxies, where $\alpha = 1$ both for NFW and Moore ($\beta = \frac{1}{2}$, for
the latter) and $\beta = 1$ both for NFW and Burkert ($\alpha = 2$, for the
latter).  Indeed, it follows from eq(\ref{EP dyn eq approxim vectorial quatro})
and assuming both conditions $\ddot{r} > 0$ and $d(\ddot{r}\,r^{2}/G)/dr \geq 0$
that the quantity $\ddot{r}\,r^{2}/G$ behaves like an additional contribution
which mimics a mass density profile as if the orbits were
circular (Mbelek 1997).  Further, $m_{eff}(r) = m(r) \,+ \,m^{(\psi)}(r) \,+
\,G^{-1}\,\ddot{r}\,r^{2}$ will define an effective mass profile consistent with
relation (\ref{generic DM profile}) in the range $r_{min} \leq r \leq r_{max}$
such that $r_{min} = \left( \,{\rho}_{halo}(r_{*})/{\rho}_{*}
\,\right)^{1/\alpha}\,r_{*}$ and $r_{max} = \left(
\,{\rho}_{*}/{\rho}_{halo}(r_{*}) \,\right)\,r_{*}$ ; ${\rho}_{halo}(r_{*}) <
{\rho}_{*}$.

\section{Results} In the following, we limit our study to those RC for which the
radial velocity, $\dot{r}$, may be neglected with respect to the tangential
velocity, $v_{\theta}$, in the whole range $r_{min} \leq r \leq r_{max}$ where
we assume that the dark halo mass density dominates the baryonic mass density.
Furthermore, the fits to the individual RC show that $\nu$ is actually an
integer $n \geq 2$.  This is consistent with our former work (Mbelek 2001) in
the sense that combining the solution found for $\psi$ with relation (\ref{halo
density}) allows us to express the right-hand side of equation (\ref{eq psi}) as
the derivative of an effective power-law potential which varies as $( \,\psi \,-
\,{\psi}_{\infty} \,)^{2(n + 1)}$.  As can be seen in figure 1, the greater the $n$, the steeper is the curve $y = v_{\theta}/v_{\theta max}$ versus $x$ for $x
< 1$ and the flater it is for $x > 1$.  Hence, the steepest is a RC below the
turnover radius, and it should be flatest beyond.  Figure 2 shows some fits to
individual rotation curves from the samples of Rubin {\sl et al.} (Rubin et al. 1980; Rubin et al. 1985), van Albada {\sl et al.}  (van Albada et al. 1985), Lake {\sl et al.}
(Lake et al. 1990).  This is achieved by using the least-squares fit to search the
parameters $a$ and $b = \ln{J}$ that yield the maximum square, $R^{2}$, of the
correlation coefficient for the relation $\ln{(r \,v_{\theta})} = a \,r^{-1/n} +
b$.  Based on the study of a hundred spirals, $a$ is
always negative, in agreement with $\delta \psi < 0$ whereas $b$ is always
positive.  In addition, one gets approximately the following statistics :  $n =
5$ for $20$\% of the spirals, $3 \leq n \leq 6$ for almost a half of them and $3
\leq n \leq 12$ for $80$\% of them.

\begin{figure}[h] \begin{center} \begin{tabular}{c}
\psfig{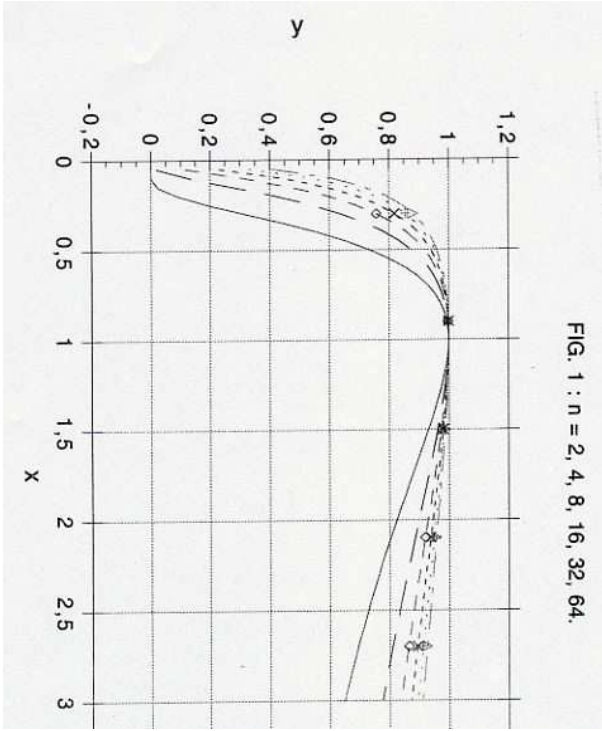} \end{tabular}
\end{center}
\caption{Generic curves $y = G_{n} (x)$ for $n = 2, 4, 8, 16, 32$ and $64$ (from bottom to top).  As
one can see, the greater $n$, the steeper is the curve below $x = 0.5$ and
the flater it is beyond.}  \label{fig:Generic curves} \end{figure}

\begin{figure}[h] \begin{center} \begin{tabular}{c}
\psfig{file=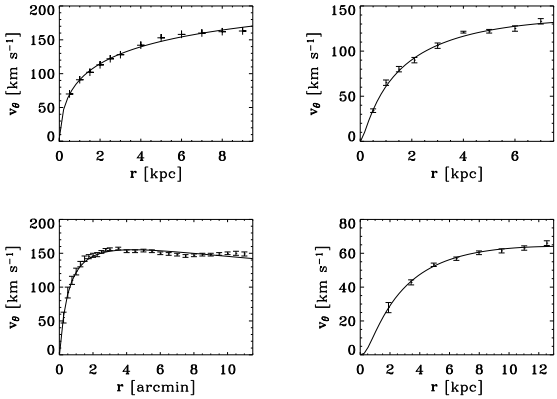,width=8.0cm} \end{tabular}
\end{center}
\caption{Rotation curve fits (from top to bottom) for NGC 4062 (Rubin et al. 1980),
NGC 1035 (Rubin et al. 1985), NGC 3198 (van Albada et al. 1985; $1$ arcmin $= 2.68$ kpc) and
DDO 170 (Lake et al. 1990) ; caption :  the crosses or error bars are the observed data
and the solid line is the best fit of the RC.}  \label{fig:Rotation curve fits} \end{figure}

\begin{figure}[h] \begin{center} \begin{tabular}{c}
\psfig{file=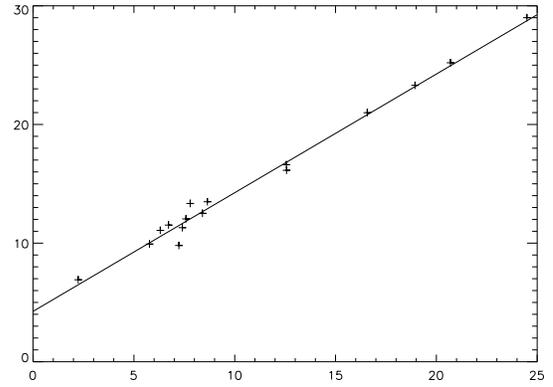,width=8.0cm} \end{tabular}
\end{center} \caption{Plot
of the coefficient $b$ versus the coefficient $\mid a \mid$ for the galaxies NGC
4062 and UGC 3691 from the sample of Rubin {\sl et al.} (Rubin et al. 1980), NGC 4419, 1035, 1325, 2742,
3067, 3593, 2639, 4378, 4448, 7606 and UGC 12810 from the sample
of Rubin {\sl et al.} (Rubin et al. 1985), NGC 3198 (van Albada et al. 1985), DDO 170 (Lake et al. 1990) and NGC
4051 (Kaneko et al. 1997).  One can see the strong linear correlation between $\mid a
\mid$ and $b$ ; the crosses are the data from the fits to individual
galaxies and the solid line is the best fit to the whole sample of a hundred
galaxies which expresses as $b = \,\mid a \mid + \,\,4.25$ with correlation
coeffient $R = 0.995$.}  \label{fig:new correlation} \end{figure}

Figure 3 shows that there is a strong correlation between both coefficients $a$
and $b$. Although displayed here for sixteen spirals only, this remains true for
the whole sample of a hundred spirals we have studied.  Expressing $r$
in $kpc$ for all galaxies, one finds\footnote{For the whole sample, $H_{0}$ is
in the range $50 \,- \,75$~km~s$^{-1}$~Mpc$^{-1}$.}  $b = \,\mid a \mid \,+
\,4.25$ with $R = 0.995$.  More generally, there are strong linear correlations
between the integer $n$ and the coefficients $a$ and $b$.  Now, the theoretical
relations between the fitting parameters $(a, b)$ and the physical
ones $(r_{0}, v_{\theta max})$ are given by $r_{0} = \left( \,\mid a \mid/n
\,\right)^{n}$ and $b = n \,+ \,\ln{( \,r_{0}\,v_{\theta max} \,)}$.  From the
latter relation, it is clear that the correlation between $b$ and $n$ is obtained
by the logarithm which alleviates the effect of the scatter in the
quantity $r_{0}\,v_{\theta max}$.  As for the correlation between $a$ and $n$,
it suggests a relation of the kind $r_{0} = \alpha \,\left( \,1 \,+ \,
\frac{\beta}{n} \,\right)^{n}$, where $\alpha \simeq 1$~kpc and $\beta$ is a
dimensionless positive constant almost independent of the given spiral galaxy.
Hence, if a natural theoretical explanation could be found for the relation
between $r_{0}$ and $n$, only one free parameter, namely $n$,
would be needed to fit the RC.  Since the latter is just an integer, it is
tempting to believe that this feature may reveal the nature of quantification underlying the galactic dark halo mass distributions.

\section{Conclusion} In this paper, we have shown based on a simple mass model
that a real scalar field minimally coupled to gravity and with matter as its source
may reproduce the RC of spiral galaxies.  This is achieved even if the
scalar field does not contribute significantly to the galactic dark matter mass.
Moreover, such a scalar field may also act as a stabilizing field.  The new
correlations that have been found open the possibility to fit the RC with only
one single free parameter which is an integer.  Thus, it seems possible
that long-range scalar fields minimally coupled to gravity play a significant
role not only at the cosmological scale (Mbelek \& Lachi\`eze-Rey 2003; del Campo \& Salgado 2003, Caresia et al. 2004) but
also within galaxies.\\

{\sl Acknowledgements.}  I thank the anonymous referee for useful
comments and suggestions that helped to improve the manuscript.


\begin{thebibliography}{}

\bibitem[1]{mbeleka} Mbelek, J.  P.  1998, Acta Cosmologica, XXIV-1, 127, and
[gr-qc/0402084]

\bibitem[2]{guzmana} Guzman, F.  S., \& Matos, T., 2000, Class.  Quant.  Grav.,
17, L9

\bibitem[3]{guzmanb} Guzman, F.  S., \& Matos, T., 2000, Phys.  Rev.  D, 62,
061301

\bibitem[4]{arbeya} Arbey, A., Lesgourgues, J., \& Salati, P.  2001, Phys.  Rev.
D, 64, 123528

\bibitem[5]{arbeyb} Arbey, A., Lesgourgues, J., \& Salati, P.  2001, Phys.  Rev.
D, 65, 083514

\bibitem[6]{ULL} Ure\~na-L\`opez, L.  A., \& Liddle, A.  R.  2002, Phys.  Rev.
D, 66, 083005

\bibitem[7]{mbelekb} Mbelek, J.  P., \& Lachi\`eze-Rey, M.  2003, A \& A, 397,
803

\bibitem[8]{narlikar} Narlikar, J.  V, \& Padmanabhan, T., 1986, Gravity, Gauge
Theory and Quantum Cosmology (D.  Reidel Publishing Company, Dordrecht), 70

\bibitem[9]{Straumann} Straumann, N., Gauge Theory and Gravitation, in Zuoz
Proceedings 2000, http://ltpth.web.psi.ch/zuoz/zuoz2000/zuoz2000proc.htm,
section 2.1

\bibitem[10]{Ciufolini} Ciufolini, I., \& Wheeler, J.  A.  1995, Gravitation and
Inertia (Princeton University Press, Princeton), 29

\bibitem[11]{Brans-Dicke} Brans, C., \& Dicke, R.  H.  1961, Phys.  Rev., 124,
925, appendix

\bibitem[12]{Wesson} Wesson, P.  S., \& Ponce de Leon, J.  1995, A \& A, 294, 1

\bibitem[13]{mbelekc} Mbelek, J.  P., Proc.  Eighth Rencontres de Blois
"Neutrinos, Dark Matter and the Universe", Stolarczyk, T., Tran Thanh Van, J.,
\& Vannucci, F.  (eds.), 391

\bibitem[14]{bournaud} Bournaud, F., Duc, P.  -A., \& Masset, F.  2003, A \& A,
411, L469

\bibitem[15]{rubina} Rubin, V.  C., Ford, W.  K.  Jr., \& Thonnard, N.  1980,
ApJ, 238, 471

\bibitem[16]{rubinb} Rubin, V.  C., Burstein, D., Ford W.  K.  Jr., \& Thonnard,
N.  1985, ApJ, 289, 81

\bibitem[17]{van albada} van Albada, T.  S., Bahcall, J.  N., Begeman, K., \&
Sancisi, R.  1985, ApJ, 295, 305

\bibitem[18]{lake} Lake, G.  L., Schommer, R.  A., \& van Gorkom, J.  H.  1990,
AJ, 99, 547

\bibitem[19]{kaneko} Kaneko, N., Aoki, K., Kosugi, G., et al.  1997, AJ, 114, 94


\bibitem[20]{del Campo} del Campo, S, \& Salgado, P.  2003, Class.  Quant.
Grav., 20, 4331

\bibitem[21]{Caresia} Caresia, P., Matarrese, S., \& Moscardini, L.  2004, ApJ.,
605, 21

\end{thebibliography}
\end{document}